\documentclass[aps,prb,amsmath,amssymb,superscriptaddress,floatfix,showpacs,twocolumn]{revtex4}

\usepackage{graphicx} 

\newcommand{\ev}[1]
{\langle #1 \rangle }

\usepackage{hyperref}

\begin{document}

\title{Zero and finite temperature mean field study of magnetic field induced electric polarization in Ba$_2$CoGe$_2$O$_7$}

\author{Judit Romh\'anyi}
\affiliation{Research Institute  for  Solid State  Physics and Optics, H--1525 Budapest, P.O.B.~49, Hungary}
\affiliation{Department of Physics, Budapest University of Technology and Economics, H--1111 Budapest, Budafoki \'ut 8, Hungary}

\author{Mikl\'os Lajk\'o}
\affiliation{Research Institute  for  Solid State  Physics and Optics, H--1525 Budapest, P.O.B.~49, Hungary}
\affiliation{Department of Physics, Budapest University of Technology and Economics, H--1111 Budapest, Budafoki \'ut 8, Hungary}

\author{Karlo Penc}
\affiliation{Research Institute  for  Solid State  Physics and Optics, H--1525 Budapest, P.O.B.~49, Hungary}

\date{\today}

\begin{abstract}   
We investigate the spin-induced polarization in the multiferroic compound Ba$_2$CuGe$_2$O$_7$ using variarional and finite temperature mean field approaches. The compound is described by a spin--3/2 Heisenberg model extended with easy plane anisotropy and Dzyaloshinskii-Moriya (DM) interaction. Applying magnetic field parallel to the $[110]$ axis, three phases can be distinguished: (i)  At high magnetic field we find a partially magnetized phase with spins parallel to the fields and uniform polarization; (ii) Below a critical field the ground state is a twofold degenerate canted antiferromagnet, where the degeneracy can be lifted by a finite DM interaction; (iii) At zero field a $U(1)$ symmetry breaking phase takes place, exhibiting a Goldstone-mode. To reproduce the magnetization and polarization measurements reported in Murakawa {\it et al.} [Phys. Rev. Lett. {\bf 105}, 137202 (2010)], we introduce an additional term in the Hamiltonian that couples the polarizations on neighboring tetrahedra. This results in the appearance of a canted ferrimagnetic phase for $h \lesssim 1 T$, characterized by a finite staggered polarization, as well as by a finite magnetization along the ${[\bar{1},1,0]}$ axis that leads to torque anomalies. 
\end{abstract}
\pacs{
75.85.+t    
75.30.Gw  
75.10.-b	
}
\maketitle

\section{Introduction}

In the traditional sense multiferroic materials are characterised by the coexistence of ferroelectricity and ferromagnetism.\cite{Fiebig2005,Mostovoy2007,Arima2011}
In these ``proper'' multiferroics the magnetoelectric effect is quite small, since the   simultaneous presence of the two order is rather difficult due to the fact that they break the inversion and time reversal symmetry in a different way:  while the ferroelectric order breaks space inversion symmetry and is invariant under time reversal, the magnetic order behaves just in the opposite way. In addition, the concurrent appearance of these orderings does not necessarily mean that the magnetic and electric dipoles are strongly coupled to each-other. 

After almost fifty years of research in the field, the discovery of the giant magnetoelectric response in TbMnO$_3$\cite{Kimura2003Nat} has launched a new concept: spin driven ferroelectricity. In the last decade --  along with experimental realizations -- many theoretical explanations have been proposed as the source of electric polarization in magnetically ordered materials. Electric polarization induced by cycloidal spin order was explained through  spin chirality\cite{Katsura2005} or inverse Dzyaloshinskii-Moriya mechanism\cite{Sergienko2006} and was found experimentally in many materials, including TbMnO$_3$,\cite{Kimura2003Nat} Ni$_3$V$_2$O$_8$,\cite{Lawes2005} CuFeO$_2$,\cite{Kimura2006} MnWO$_4$,\cite{Taniguchi2006} CoCr$_2$O$_4$,\cite{Yamasaki2006} LiCu$_2$O$_2$\cite{Park2007} and CuO\cite{Kimura2008}, just to mention a few. Exchange striction can also lead to electric polarization, as was shown in the case of the perovskite $R$MnO$_3$ materials, with $R$ being a rare earth ion.\cite{Mochizuki2010} The aforementioned  mechanisms may induce polarization jointly, as predicted in the case of $R$Mn$_2$O$_5$ materials.\cite{Chapon2006,Noda2008,Fukunaga2009} In all these cases the mechanism involves  two spins on a bond. 

If the spin is located on a site that breaks the inversion symmetry, the polarization can be induced by a single spin. The origin of this process is the spin dependent metal-ligand hybridization arising from spin-orbit coupling.\cite{Jia2006,Jia2007} 
Murakawa and collaborators proposed that this mechanism explains the induced ferroelectric polarization in Ba$_2$CoGe$_2$O$_7$.\cite{Murakawa2010}

As a result of strong easy-plane anisotropy, below $T_N=6.7$ K the $S=3/2$ moments order into a canted antiferromagnetic pattern that is confined in the Co--plane.\cite{Zheludev2003}
This canted planar antiferromagnetic phase is a multiferroic phase in which magnetoelectric behavior has been observed.  Ascribed to the symmetry properties of Ba$_2$CoGe$_2$O$_7$ the sum over the vector spin chirality ${\bf S_i}\times{\bf S_j}$ vanishes and the exchange interaction ${\bf S_i}\cdot{\bf S_j}$ is uniform for all bonds. Hence the induced polarization cannot be explained by the concept of spin current or exchange striction. Spin dependent hybridization mechanism was suggested as a solution, according to which the local polarization takes the form of ${\bf P}\propto\sum^{4}_{i=1}({\bf S}\cdot{\bf e_i})^2{\bf e_i}$, where ${\bf e_i}$ vectors point from the Co ions toward the surrounding four O ions.\cite{Murakawa2010} This model intrinsically possesses the symmetries of the lattice and gives the same result for the magnetization dependence of the polarization vector as a thorough symmetry analysis (e.g. in Ref.~[\onlinecite{Perez-Mato2010}]) would. It recovers the sinusoidal response of electric polarization to the rotating magnetic field and describes the nature of induced polarization in magnetic field qualitatively well.\cite{Murakawa2010}

In this paper we investigate the magnetic field dependence of induced electric polarization in Ba$_2$CoGe$_2$O$_7$. 
The paper is structured as follows:  we start with a short symmetry analysis in Sec. \ref{sec:symm} to deduce the Hamiltonian and the spin dependent expression of the polarization.  In Sec. \ref{sec:T0} we study the field dependence of polarization at zero temperature variationally, using a site factorized wave function. Two cases are considered, when the external field is applied along the $[110]$ and $[100]$ axes. We investigate the effect of DM interaction and an additional antiferro polarization term in detail. In Sec. \ref{sec:finiteT} we perform finite temperature mean field approximation to calculate the polarization and magnetization for both field settings. We qualitatively compare our $h||[110]$ results to the experimental findings of Ref.~[\onlinecite{Murakawa2010}] and give predictions for the induced polarization with regard the $h||[100]$ field setting.

\section{Symmetry considerations}\label{sec:symm}

\begin{figure}[h!]
\begin{center}
\includegraphics[width=7cm]{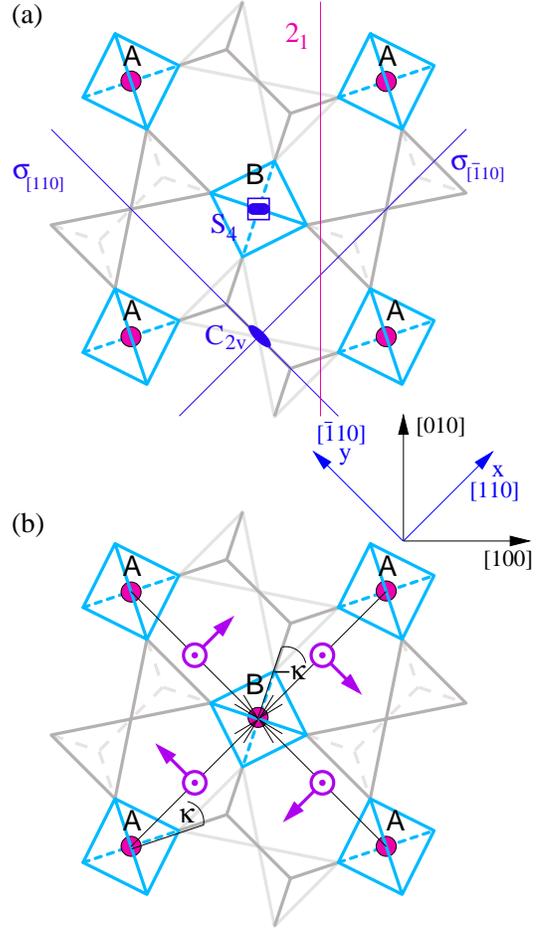}
\caption{A schematic figure of Ba$_2$CoGe$_2$O$_7$. (a) the point group of this material can be constructed from the groups $\mathcal{S}_4$ and $\mathcal{C}_{2v}$ acting on the Co sites and in the middle of four sites, respectively. The mirror planes are labeled by their normal vector. (b) due to the tilted tetrahedral environment the unit cell contains two Co$^{2+}$ ions that we denote by $A$ and $B$. The symmetry allowed DM vectors -- with respect to the order of involved sites (black arrows) -- are indicated by purple arrows. At the $\Gamma$ point the in--plane components of DM vector cancel each-other thus in our case it is sufficient to take only the out--of--plane components into account.}
\label{fig:BCGO}
\end{center}
\end{figure}

A schematic figure of Ba$_2$CoGe$_2$O$_7$ is shown in Fig. \ref{fig:BCGO}. The
Ba$_2$CoGe$_2$O$_7$ has a tetragonal non-centrosymmetric structure  characterized by the space group $P\overline{4}2_1m$. The neighboring CoO$_4$ tetrahedra are tilted from the $[110]$ crystallographic direction by the angles $\pm\kappa$, and due to the differently oriented tetrahedral environments the unit cell contains two Co$^{2+}$ ions with $S=3/2$. 

The point group of the lattice is isomorphic to the group $\mathcal{D}_{2d}$ that consists of the groups  $\mathcal{S}_4=\{E,S_4,C_2,S^3_4\}$, with the rotation axis located on the Co sites, and of $\mathcal{C}_{2v}=\{E,C_2,\sigma_{[\overline{1}10]},\sigma_{[110]}\}$ in the center of the four Co sites, as shown in Fig. \ref{fig:BCGO}(a). 
The symmetry properties of the lattice determine the terms that can be included in the Hamiltonian, i.e. the terms that transform as the fully symmetric irreducible representation A$_1$ of $\mathcal{D}_{2d}$. When classifying the operators we assume that the magnetic order is a two-sublattice order, i.e. the magnetic unit cell corresponds to that of the lattice. Our Hamiltonian has the form of
\begin{eqnarray}
\mathcal{H}&=&J\sum_{(i,j)}\left(\hat S^x_i \hat S^x_j+\hat S^y_i \hat S^y_j\right)+J_z \sum_{(i,j)}\hat S^z_i \hat S^z_j+\Lambda\sum_{i}(\hat S^z_i)^2\nonumber\\
&{}&- g {\bf h} \sum_{i}{\bf \hat S}_i +D_z\sum_{(i \in A,j \in B)}\left(\hat S^x_i \hat S^y_j-\hat S^y_i \hat S^x_j\right)\;,
\label{eq:Hamiltonian}
\end{eqnarray}
where $(i,j)$ denotes pairs of nearest neighbor sites. 
The orientation of the DM vector (see Fig.~\ref{fig:BCGO}(b)) is fixed by the point group of the unit cell. 
The lattice symmetries allow for other terms, too, such as the g-tensor anisotropy:
\begin{eqnarray}
 + g_s \sum_{i\in A} \left(h_x S^y_i - h_y S^x_i \right) + g_s \sum_{i\in B}
\left( h_y S^x_i - g_s h_x S^y_i \right) \;
\end{eqnarray}
with $h_x=h_{[110]}$ and $h_y=h_{[\overline{1}10]}$. Although, we found that it has no significant effect on the field dependence of the polarization, therefore we did not include it in our model.

As mentioned before, the polarization P and the spin behave in different ways under inversion ($\mathcal{I}$) and time reversal ($\mathcal{T}$) operations: $\mathcal{I} P=-P$ and $\mathcal{T} P=P$, while $\mathcal{I} S=S$ and $\mathcal{T} S=-S$. 
For there is no inversion center in Ba$_2$CoGe$_2$O$_7$, we only have to account for $\mathcal{T}$, and it follows that the polarization can be coupled linearly to operators that are even-order in spin and consequently are invariant under time reversal. With respect to the lattice symmetry, the $z$-component of polarization vector -- or in fact any polar vector -- transforms as the irreducible representation $B_2$ of $\mathcal{D}_{2d}$, while the in-plane components $(p^x,p^y)$ belong to the two dimensional representation $E$. Classifying the second-order spin operators according to lattice symmetry we find that, e.g. $(\hat{S}^y_{j})^2-(\hat{S}^x_{j})^2$ and $\hat{S}^x_{j}\hat{S}^y_{j}+\hat{S}^y_{j}\hat{S}^x_{j}$ transform as the irreducible representation $B_2$ as well, therefore they can be coupled to the $z$-component of polarization. Similar logic leads to the form of the operators $\hat{P}^{x}$ and $\hat{P}^{y}$:
\begin{eqnarray}
\hat{P}^{x}_{j} &\propto&
 -\cos2\kappa_j\left(\hat{S}^x_{j}\hat{S}^z_{j}+\hat{S}^z_{j}\hat{S}^x_{j}\right)-\sin2\kappa_j\left(\hat{S}^y_{j}\hat{S}^z_{j}+\hat{S}^z_{j}\hat{S}^y_{j}\right)
 \nonumber\\
\hat{P}^{y}_{j}&\propto&
\cos2\kappa_j\left(\hat{S}^y_{j}\hat{S}^z_{j}+\hat{S}^z_{j}\hat{S}^y_{j}\right)-\sin2\kappa_j\left(\hat{S}^x_{j}\hat{S}^z_{j}+\hat{S}^z_{j}\hat{S}^x_{j}\right)
\nonumber\\
\hat{P}^{z}_{j}&\propto&
\cos2\kappa_j\left((\hat{S}^y_{j})^2-(\hat{S}^x_{j})^2\right)-\sin2\kappa_j\left(\hat{S}^x_{j}\hat{S}^y_{j}+\hat{S}^y_{j}\hat{S}^x_{j}\right)
\label{eq:pol}
\end{eqnarray}
where $j$ belongs to either sublattice $A$, or $B$. The different orientation of the tetrahedra is accounted for by choosing $\kappa_{j\in A}=\kappa$ and $\kappa_{j\in B}=-\kappa$. Note that the operators defined in (\ref{eq:pol}) are actually quadrupole operators.

\section{Zero temperature variational approach}
\label{sec:T0}

Based on the neutron scattering measurements of Ref.~[\onlinecite{Zheludev2003}] that suggested a two-sublattice canted antiferromagnet as the ground state, we chose a site factorized variational wave function:
\begin{equation}
|\Psi\rangle = \prod_{i\in A}\prod_{j\in B} |\psi_{i}\rangle |\psi_j\rangle \;,
\label{eqn:var-ansatz}
\end{equation}
with
\begin{equation}
| \psi_A \rangle \propto u_{0} |\frac{3}{2}\rangle+e^{i\xi_1}u_{1}|\frac{1}{2}\rangle+e^{i\xi_2}u_2|-\frac{1}{2}\rangle+e^{i\xi_2}u_3|-\frac{3}{2}\rangle 
\label{eq:varpsiABA}
\end{equation}
and 
\begin{equation}
|\psi_B \rangle \propto v_{0} |\frac{3}{2}\rangle+e^{i\vartheta_1}v_{1}|\frac{1}{2}\rangle+e^{i\vartheta_2}v_2|-\frac{1}{2}\rangle+e^{i\vartheta_3}v_3|-\frac{3}{2}\rangle \;
\label{eq:varpsiABB}
\end{equation}
where the $|3/2\rangle$, $|1/2\rangle$, \dots denote the $|S^z\rangle$ of the spin--3/2 on the A and  B sites. There are 6 independent  variational parameters for 
$| \psi_A \rangle$ and another 6 for $| \psi_B \rangle$ to be determined by minimizing the ground state energy
\begin{equation}
  E = \frac{\langle \Psi | \mathcal{H} | \Psi \rangle}{\langle \Psi | \Psi \rangle} \;.
  \label{eqn:var-energy}
\end{equation}
For simplicity we assume that there is no exchange anisotropy: $J=J_z$ -- as in Ref.~[\onlinecite{Miyahara2011}]. We study the effect of the DM interaction in detail. 

\subsection{Mapping between the cases $h||[110]$ and $h||[100]$} \label{sec:mapping}

\begin{figure}[h!]
\begin{center}
\includegraphics[width=8cm]{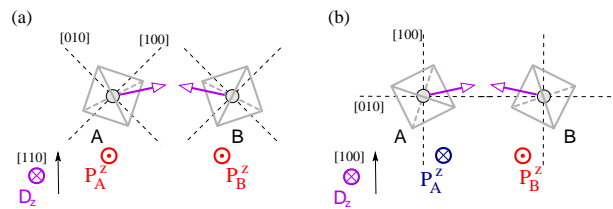}
\caption{Relation between the cases $h||[110]$ and $h||[100]$. In our notation $P^z_j$ takes maximum value for a spin aligned parallel to the lower edge (dashed line) and minimum value along the upper (solid) edge. (a)ÊSchematic spin configuration for $h||[110]$ and $D_z<0$. The spin induced polarization is the same on the two sublattices for the spins are canted by the same angle from the same (lower) edge of the tetrahedra. (b) Spin configuration for $h||[100]$ and $D_z<0$ after rotating the basis by the angle $\pi/4$. Note that due to the rotation of the tetrahedral environment the induced polarization changes. The spins of the two subblatices are now canted from the different edges, consequently the induced polarization on them have different sign.}
\label{fig:basis_tr}
\end{center}
\end{figure}

In the following sections we compare two cases: when the field is applied along the axes $[110]$ and $[100]$. Whereas regarding the magnetization it is natural to consider $m_{[110]}$ and its rotated counterpart $m_{[100]}$ as the relevant magnetic order parameters, it is not that easily perceived which component of the polarization will characterize the electric order. 
Forgetting about the polarization, the two kinds of field settings can be mapped to each other by a  $\pi/4$-rotation of the local basis. We introduce the new spin operators so that $\tilde{\hat{S}}^x$ is the spin component parallel to the applied field $h_{[100]}$: $\tilde{\hat{S}}^x_j=1/\sqrt{2}(\hat{S}^x_j-\hat{S}^y_j)$ and $\tilde{\hat{S}}^y_j=1/\sqrt{2}(\hat{S}^y_j+\hat{S}^x_j)$. In the case $h||[1,0,0] $, the Hamiltonian in the new basis has exactly the same form as (\ref{eq:Hamiltonian}) when the field is parallel to $[1,1,0]$. The polarization operators, on the other hand, change. Considering the $z$-component of $\hat{P}$, in the new basis it will have the form of
\begin{eqnarray}
\tilde{\hat{P}}^{z}_{j}&=&\cos2\tilde{\kappa}_j\left((\tilde{\hat{S}}^y_{j})^2-(\tilde{\hat{S}}^x_{j})^2\right)-\sin2\tilde{\kappa}_j\left(\tilde{\hat{S}}^x_{j}\tilde{\hat{S}}^y_{j}+\tilde{\hat{S}}^y_{j}\tilde{\hat{S}}^x_{j}\right)
\end{eqnarray}
with $\tilde{\kappa}=\kappa+\pi/4$.
In order to an easier understanding of the relation between induced polarization in the two basis we sketched the effect of local basis transformation in Fig. \ref{fig:basis_tr}. As it was pointed out in Ref.~[\onlinecite{Murakawa2010}], the induced polarization has extremum when the spin is aligned with either the upper or the lower edge of the surrounding tetrahedra. (The lower and upper edges are those parallel to the Co-plane.) If $P^z_j$ is maximal for a given spin orientation, the $\pi/2$-rotation of the spin will cause the polarization to change sign. When the spin lays halfway between these two positions, the polarization is zero.
Applying the field in the direction $[110]$, the spins on the two sublattices are tilted (by the same angle) from the same symmetry plane of the tetrahedra, therefore the induced polarization on $A$ and $B$ has the same sign (and magnitude). However a $\pi/4$ rotation of the local basis rotates the tetrahedral environment in a way that the spins on $A$ and $B$ are now tilted from the different edges, and as a result the induced polarizations have different sign on the two sublattices. (see Fig. \ref{fig:basis_tr})

According to this argument, when the field is parallel to $[110]$ the relevant order parameter is the total polarization $\hat{P}^z=\hat{P}^z_A+\hat{P}^z_B$, whereas for $h||[100]$ the staggered component $\hat{P}^{\text{st}}_z=\hat{P}^z_B-\hat{P}^z_A$ will give a finite expectation value. We shall note, that the difference between the two cases is related to the lattice symmetries as well. While along the $[110]$ direction the symmetry operation that connects site $A$ with site $B$ is a $\sigma_{[110]}$ mirror plane, along the $[100]$ direction the sites can be transformed into each-other by a $2_1$ screw axis. (see Fig. \ref{fig:BCGO} (a))

\subsection{$D_z=0$}

Let us begin our investigations with the case $D_z=0$. In zero magnetic field the variational ground state of (\ref{eq:Hamiltonian}) is a planar antiferromagnet (superfluid) that breaks rotational symmetry $U(1)$, and in which the staggered in--plane magnetic order parameter $O_{U(1)}=\frac{1}{2}|{\bf \hat{S}}^{\bot}_A-{\bf \hat{S}}^{\bot}_B|$ has finite expectation value, where ${\bf \hat{S}}^{\bot}_j=(\hat{S}^x_j,\hat{S}^y_j)$.\cite{Romhanyi2011} The planarity is the consequence of the easy-plane single ion anisotropy and the $U(1)$ symmetry breaking is related to the fact that in the absence of an in-plane magnetic field the Hamiltonian commutes with total $S^z$.
The ground state wave function of site $A$ can be expressed as
\begin{eqnarray}  
|\Psi_A\rangle &=& e^{-i \varphi_A \hat S^z_A} |\Psi^{\text{SF}}_A\rangle
\label{eq:planar_AB}
\end{eqnarray}
with a single variational parameter $\eta_A$:
\begin{equation}
|\Psi^{\text{SF}}_A\rangle = \frac{|\frac{3}{2}\rangle + |-\frac{3}{2}\rangle 
+ \sqrt{3}\eta_A \left(|\frac{1}{2}\rangle +  |-\frac{1}{2}\rangle \right) }{\sqrt{6 \eta_A^2 + 2}} .
\label{eq:planar_grst}
\end{equation}
and a similar expression stands for $|\Psi_B\rangle$ with parameters $\varphi_B$ and $\eta_B$. $\varphi_A$ and $\varphi_B$ determine the angles of the spins with respect to the $[110]$ direction. Furthermore we find that $\eta_A=\eta_B=\eta\geq 1$ for the whole magnetization process. In zero field $\varphi_A=\varphi$ and $\varphi_B=\varphi-\pi$, i.e. the spins are antiparallel.\cite{Romhanyi2011} The $U(1)$ symmetry breaking is manifested in the fact that $\varphi$ can take arbitrary value.

In a finite magnetic field $\varphi$ is not arbitrary any longer and the canted spins turn so that the total magnetization points into the direction of the field. This two sublattice canted order is twofold degenerate because, in the absence of the DM term, the Hamiltonian is invariant under the exchange of sites $A$ and $B$. With increasing field the angle $\delta\varphi=\varphi_A-\varphi_B$ between the spins of the sublattices decreases from $\delta\varphi=\pi$ to $\delta\varphi=0$, while the length of the spins is unchanged ($\eta$ is constant). 

At a critical value ($h_c$) the two spins become aligned: $\varphi_A=\varphi_B=0$ for the field parallel to $[110]$ and $\varphi_A=\varphi_B=-\pi/4$ for $h||[100]$. The two sublattice order vanishes and a uniform phase appears. The critical field, however, is not equal to the saturation field, the spins are not fully magnetized yet. With the further increase of the field the magnetization increases slowly as the spins reach their full length (i.e. $\eta$ decreases to $\eta=1$). 

 The polarization and magnetization for $D_z=0$ in magnetic fields along the axes $[110]$ and $[100]$ are plotted with black solid lines in Fig. \ref{fig:DM_effect}(c) and \ref{fig:DM_effect_h100}(c), respectively. The spin configurations of the twofold degenerate canted order are shown in panel (a) and (b) of Figs. \ref{fig:DM_effect} and \ref{fig:DM_effect_h100}. In both cases we indicated the canting angle of the spins with respect to the applied field. Conveniently, in the case of $h||[110]$ this canting angle is $\varphi_A$ and $\varphi_B$ -- following from the definition of the ground state (\ref{eq:planar_AB}). However, when $h||[100]$ we need to introduce the canting angles $\tilde{\varphi}_A$ and $\tilde{\varphi}_B$, to measure the angle from the $[100]$-axis.  
Actually $\tilde{\varphi}_A$ and $\tilde{\varphi}_B$ play the same role in the rotated basis introduced in Sec. \ref{sec:mapping} (when $\tilde{\hat{S}}^x$ is the spin component along $[100]$) as $\varphi_A$ and $\varphi_B$ in the original basis. Furthermore one could express the ground state (\ref{eq:planar_AB}) in the rotated basis, using  $\tilde{\varphi}_{A/B}=\varphi_{A/B}+\pi/4$.

\subsection{Finite $D_z$}

When the DM interaction is finite, the angle between the spins of the sublattices $A$ and $B$ varies from $\pi$, ($\varphi_A-\varphi_B\neq\pi$), and the spins become canted even in zero magnetic field, with a canting angle depending on the magnitude of $D_z$. 
Nonetheless, this canted antiferromagnetic state breaks the $U(1)$ symmetry too, for $D_z$ controls only the direction of the spins on different sublattices relative to each other, and $S^z$ remains a good quantum number ($[\mathcal{H},S^z]=0$) as long as there is no field applied in the plane.
A finite magnetic field lifts the $U(1)$-degeneracy and we enter into the canted antiferromagnetic phase, where the canting angle depends on $D_z$ and the field. 
The finite DM coupling -- being sensitive to the exchange of $A$ and $B$ -- lifts the twofold degeneracy of the canted phase and depending on its sign either of these two states is preferred. 
(A sign difference in ${\bf D}\cdot ({\bf \hat{S}}_A\times {\bf \hat{S}}_B)$ is equivalent to the exchange of sites $A$ and $B$). Similarly to the case $D_z=0$, the variational parameters are $\eta_A=\eta_B=\eta$, and below the critical field $\eta$ -- consequently the spin length -- is constant, as opposed to $h>h_c$, when the spins are partially magnetized and their length increases with the field.
Comparing our results to the measurement of Ref.~[\onlinecite{Murakawa2010}]  in the case of $h||[110]$, we find that $D_z<0$ is the appropriate choice.

\begin{figure}[h]
\begin{center}
\includegraphics[width=8cm]{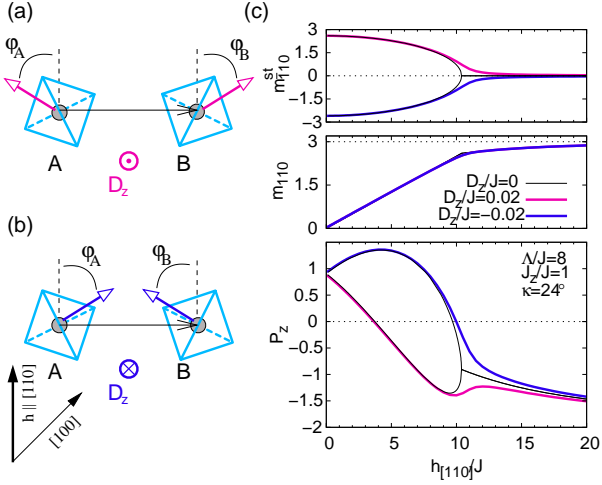}
\caption{(a),(b) Schematic figure of ground states in canted antiferromagnetic phase when $h||[110]$. For $D_z=0$ (a) and (b) are degenerate. A finite DM coupling lifts the twofold degeneracy, the ground state configuration for $D_z>0$ is shown in (a) while (b) is selected when $D_z<0$. (c) The polarization $P_z$, the magnetization $m_{[110]}$ and the staggered magnetization $m^{\text{st}}_{[\overline{1}10]}$ is shown. Black line indicates the twofold degeneracy of the $D_z=0$ case below the partially magnetized phase. The colored lines correspond to the $D_z>0$ and $D_z<0$ cases in accordance with the coloring of spin states in (a) and (b).}
\label{fig:DM_effect}
\end{center}
\end{figure}

\begin{figure}[h]
\begin{center}
\includegraphics[width=8cm]{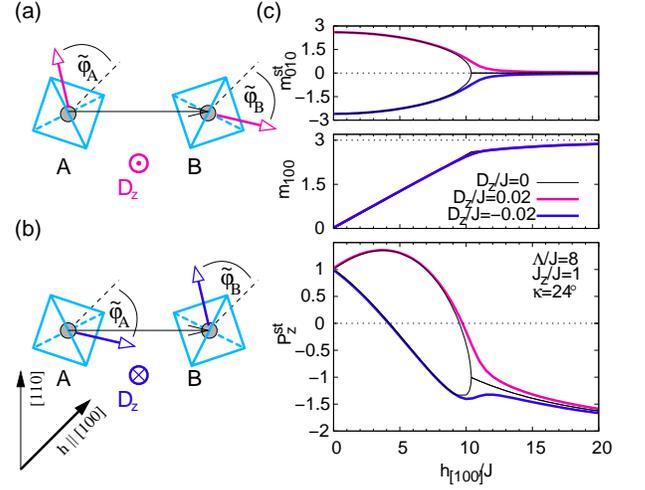}
\caption{(a),(b) the canted antiferromagnetic ground state for $h||[100]$. Similarly to the case of $h||[110]$ (a) and (b) are degenerate when $D_z=0$. Whereas a finite $D_z$ lifts this degeneracy, and $D_z>0$ selects the canted state (a) while $D_z<0$ prefers the configuration of (b).  (c) When  $h||[100]$ the uniform polarization vanishes ($P_z=P^A_z+P^B_z=0$) and the staggered polarization $P^{\text{st}}_z=P^B_z-P^A_z$ gives a finite value. The coloring of the cases $D_z>0$ and $D_z<0$ follows that of the spin states used in (a) and (b).}
\label{fig:DM_effect_h100}
\end{center}
\end{figure}

A schematic figure of the ground states in the canted phase is shown for $h || [110]$ in Figs.~\ref{fig:DM_effect}(a) when $D_z>0$ and \ref{fig:DM_effect}(b) when $D_z<0$, and correspondingly in Figs. \ref{fig:DM_effect_h100}(a) and (b) for $h || [100]$.
The magnetization and the spin order induced polarization is plotted in Figs.~\ref{fig:DM_effect}(c) and \ref{fig:DM_effect_h100}(c). Based on the argument of Sec.~\ref{sec:mapping}, when the field is pointing in the direction $[110]$ the total polarization is finite, while for $h||[100]$ the staggered polarization gives non-zero expectation value. We remark that with the choice of $ \kappa = \pi/8 $, the total polarization for $h||[[1,1,0]$ and the staggered polarization for $h||[1,0,0]$ would match perfectly. A finite $D_z$ interaction chooses different branches of the electric order parameter ($P_z$ and $P^\text{st}_z$) for the two settings of magnetic field, e.g. $D_z < 0$ favors the higher branch of the total polarization for $h||[1,1,0]$ and the lower branch of the staggered polarization in the case of $h||[1,0,0]$.

\subsection{Effect of antiferro polarization term}

\begin{figure}[h]
\begin{center}
\includegraphics[width=8cm]{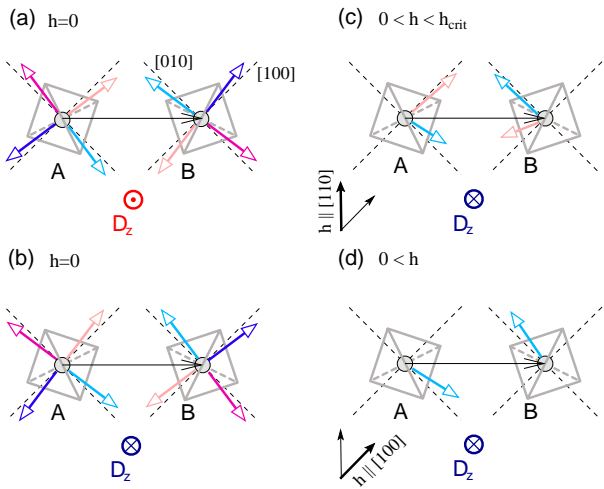}
\caption{Ground state spin configurations when the antiferro polarization is present in the Hamiltonian. (a) and (b) the fourfold degenerate ground state in zero field for different signs of $D_z$. A spin configuration -- with smaller canting angle -- shown in (a) would correspond to the case $D_z=0$, too (c) when $h||[110]$ a canted ferrimagnetic phase emerges below $h_{c2}\approx 1$ T due to the polarization term. (d) for $h||[100]$, since $K_z>0$, the canted ferrimagnetic phase is missing, and at finite field the ground state is the canted antiferromagnet similar to the case $K_z=0$. A ferromagnetic polarization term would have the opposite effect.}
\label{fig:states}
\end{center}
\end{figure}

 For higher fields the polarization curve is satisfyingly similar to the measurements, however, the low field behavior cannot be explained using only the Hamiltonian (\ref{eq:Hamiltonian}). In order to reproduce the sharp decrease of polarization below $h \approx 1$ T we add an anitiferro ($K_z>0$) polarization term to our model:
\begin{eqnarray}
\mathcal{H}_{\text{pol}}=K_z \sum_{(i,j)}\hat{P}^z_i \hat{P}^z_j\;
\label{eq:H_pol}
\end{eqnarray}
This is a kind of anisotropic biquadratic term that is allowed by the symmetry. 
Since $[\mathcal{H}_{\text{pol}},S^z] \neq 0$, the $U(1)$ symmetry is lost with this term in the Hamiltonian.
Instead, 
a fourfold degenerate ground state appears (actually this corresponds to the ground state discussed in Ref.~\onlinecite{Toledano2011}, where the order parameters in Ba$_2$CoGe$_2$O$_7$ were investigated from purely the aspect of symmetries). We start our investigation with $D_z=0$ again. The spin direction is determined by the minimization of (\ref{eq:H_pol}) and (\ref{eq:Hamiltonian}). As it was discussed in Ref.~[\onlinecite{Murakawa2010}] and Sec. \ref{sec:mapping}, the orientation of the spins relative to the surrounding tetrahedron determines the induced polarization.(see Fig. 3(a) in Ref.~[\onlinecite{Murakawa2010}]). When the spin is pointing along the lower (or upper) edge of the tetrahedron, the induced $P^z$ is maximal (or minimal). Therefore a term $\hat{P}^z_A \hat{P}^z_B$ favors a spin configuration in which the spins on sites $A$ and $B$ are parallel to different edges. (Note that this is not an orthogonal spin configuration for the tetrahedra of the different sublattices are rotated by $2\kappa\approx 48^{\circ}$ compared to each other.) 
The polarization term competes with the antiferromagnetic exchange interaction and the resulting ground state is the canted state shown in Fig. \ref{fig:states}(a). A finite magnetization points along the [100] or [010] axes in these cases.

When the DM interaction is finite this spin configuration is favored by $D_z>0$ that only changes the canting angle. However, a $D_z<0$ introduces frustration, as it prefers a reversed spin orientation: if the angle between the spins is $\varphi_A-\varphi_B$ for $D_z>0$, it is $-\varphi_A+\varphi_B$ for $D_z<0$ (see Fig. \ref{fig:states}(b)). Loosely speaking this corresponds to the exchange of sites $A$ and $B$, however with a smaller canting angle due to the competition of $K_z$ and $D_z$.

Selecting $D_z<0$ --- that recovers the experimental results --- we calculated the induced polarization for both orientations of the magnetic field. The variational and the finite temperature mean field results are shown together in Fig. \ref{fig:h110_Pz_a} and \ref{fig:h110_Pz_b} for $h||[110]$ and in Fig.  \ref{fig:h100_Pz} when $h||[100]$.
In the former case we observed a new phase below $h_{c2}\approx 1$ T. 
This intervening phase is twofold degenerate. Following from the different spin length on the sublattices $A$ and $B$ a suitable order parameter is $|{\bf S}_A|-|{\bf S}_B|$, therefore we refer to this phase as a canted ferrimagnetic phase. The ground state can still be written as in (\ref{eq:planar_grst}), but here $\eta_A\neq\eta_B$. The angle between the spins on sublattices $A$ and $B$ is determined by the collective effect of $K_z$, $D_z$ and $h_{[110]}$. A schematic spin configuration of this phase is shown in Fig. \ref{fig:states}(c).
When $h>h_{c2}\approx1$ T we enter the canted antiferromagnetic phase that was characteristic in the case $K_z=0$ as well (see Fig. \ref{fig:DM_effect} (a)). When the field exceeds $h_c\approx 13$ T, the partially magnetized uniform phase emerges.
In the rotated field setting ($h||[100]$) the intervening phase is missing, and only three phases are observed. The finite field lifts the fourfold degeneracy of the $h=0$ ground state and we enter the non-degenerate two sublattice canted antiferromagnetic order shown in Figs.~\ref{fig:states}(d) or \ref{fig:DM_effect_h100}(b). At the critical field the canted antiferromagnet is replaced by the uniform phase. 
We shall note that a ferro polarization coupling ($K_z<0$) would reverse the situation, and it would cause the emergence of the canted ferrimagnetic phase when the field is $h||[100]$, while for $h||[110]$ this phase would not be present.

\section{Finite temperature mean field theory and comparison with the experiment}\label{sec:finiteT}

Assuming site factorized solution for the finite temperature mean field calculations as well, we can write the Hamiltonian as $\mathcal{H}_\text{MF} = \mathcal{H}_A+\mathcal{H}_B $, where
\begin{eqnarray}\label{eqn:Hmf}
\mathcal{H}_A &=& \phantom{+} 4 J \mathbf{S}_A \ev{\mathbf{S}_B} + 4 D_z \left( S_A^x \ev{S_B^y} - S_A^y \ev{S_B^x}\right)\nonumber \\
&{}& + 4 K_z P_A^z \ev{P_B^z} + \Lambda \left( S_A^z \right)^2  - g \mathbf{h} \cdot \mathbf{S}_A ,
\end{eqnarray} 
and we obtain $\mathcal{H}_B$ by exchanging the sublattices A and B and the sign of $D_z$. $\mathcal{H}_A$ and $\mathcal{H}_B$ acts only on sublattice A and B, respectively and the factor 4 corresponds to the coordination number. The  $\ev{\mathbf{S}_B}$ and $\ev{P^z_B}$ denote the thermodynamical average given as
 \begin{eqnarray}
 \ev{\mathbf{S}_B} & = &
  \frac{\text{Tr} \big( \mathbf{S}_B e^{-\beta \mathcal{H}_B \left( \ev{\mathbf{S}_A}, \ev{P_A^z} \right) } \big)}{ \text{Tr} \, e^{-\beta \mathcal{H}_B  \left( \ev{\mathbf{S}_A}, \ev{P_A^z} \right)}}  ,
 \nonumber \\
 \ev{P_B^z} & = & \frac{\text{Tr} \big( P_B^z  e^{-\beta \mathcal{H}_B \left( \ev{\mathbf{S}_A}, \ev{P_A^z} \right)} \big)}{\text{Tr} \, e^{-\beta \mathcal{H} _B  \left( \ev{\mathbf{S}_A}, \ev{P_A^z} \right)} }.
 \end{eqnarray}
 with $\beta=1/T$. A similar expressions stand for $\ev{\mathbf{S}_A}$ and $\ev{P^z_A}$ with the interchange of $A$ and $B$. The self-consistent solutions of this set of equations, that constitute the mean-field approximation, is obtained by iteration. 

\begin{figure}[h]
\begin{center}
\includegraphics[width=8cm]{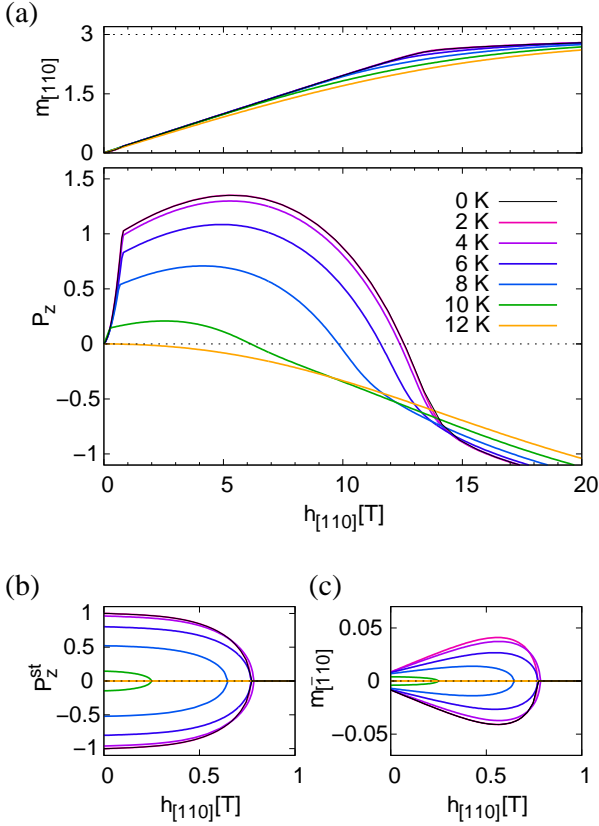}
\caption{(a) behavior of magnetization and polarization in external field along $[110]$ at various temperature values. (b), (c) for fields $h\lessapprox 1$ T we observe the canted ferrimagnetic phase that is the consequence of the antiferro polarization term (\ref{eq:H_pol}). In this phase the ground state is twofold degenerate as seen in the nature of $P^{\text{st}}_z$ and $m_{[\overline{1}10]}$.($J=J_z=1.885$ K, $\Lambda=15.08$ K, $D_z=-0.02$ K, $K_z=0.01$ K and $g=2.2$)}
\label{fig:h110_Pz_a}
\end{center}
\end{figure}

\begin{figure}[h]
\begin{center}
\includegraphics[width=8cm]{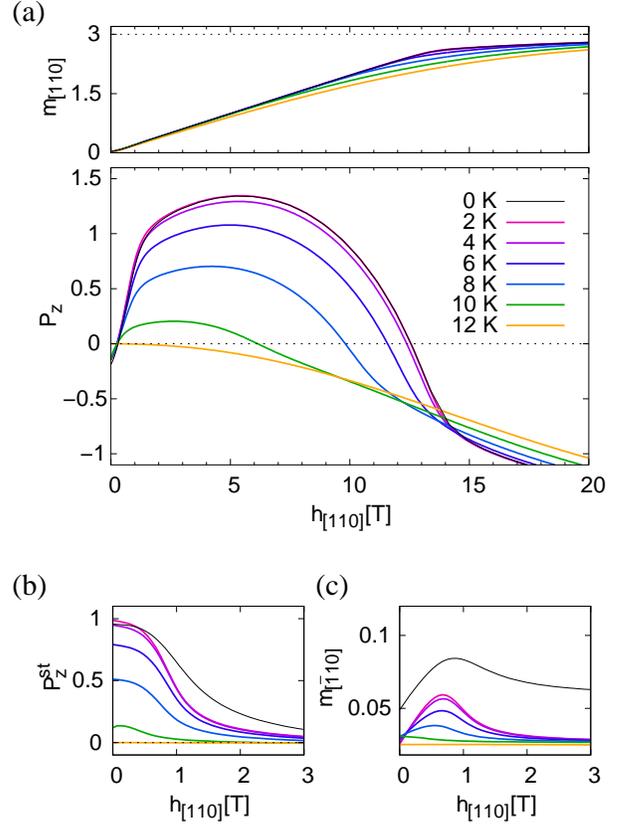}
\caption{As a possible solution to the differences in our results and the measurements, we applied a small magnetic field perpendicular to $[110]$, e.g. $h_\varepsilon||[\overline{1}10]$ and found that the characteristics of the experimental results are qualitatively well reproduced.($J=J_z=1.885$ K, $\Lambda=15.08$ K, $D_z=-0.02$ K, $K_z=0.01$ K and $g=2.2$)}
\label{fig:h110_Pz_b}
\end{center}
\end{figure}

 In our finite T calculations we took a realistic parameter setting: $J=J_z=1.885$ K, $\Lambda=15.08$ K according to Ref.~[\onlinecite{Miyahara2011}], $g=2.2$ and the values $D_z$ and $K_z$ were set to $-0.02$ K and $0.01$ K respectively. 
The mean field result for T = 2 K is essentially undistinguishable from the T=0 variational calculation.
For the field applied parallel to $[110]$, we find that below $h_{c2}\approx 1$ T the polarization drops to zero and in this region there is a finite expectation value of $m_{[\overline{1}10]}$ and $P^{\text{st}}_z$ corresponding to the zero temperature canted ferrimagnetic phase. The fact that these functions can take positive and negative values reflects the twofold degeneracy of the ground state. At higher field we enter the canted antiferromagnetic region and at $h_c$ there is a continuous phase transition to the partially magnetized uniform phase where the spins are aligned with the external field. With further increase of $h_{[110]}$ the spins grow to reach their full length. The finite temperature results for magnetic field $h||[1,1,0]$ are summarized in Fig. \ref{fig:h110_Pz_a}. Taking strictly the Hamiltonian (\ref{eq:Hamiltonian}) extended with the polarization term (\ref{eq:H_pol}) we can recover the main characteristics of the experimental findings in Ref.~[\onlinecite{Murakawa2010}]. The polarization drops sharply below $1$ T and it changes drastically with increasing temperature, whereas the magnetization curve is almost unchanged. Torque measurements would be conclusive to our studies, and could reveal if there is an orthogonal component of magnetization at low fields ($h<h_{c2}$).
Nonetheless, there are yet some properties to account for. The polarization curve in the experiments starts from negative value at zero field, while in our model --- due to the antiferro polarization term --- $P_z=0$ when $h=0$. Furthermore around $h_{c2}$ the shape of the polarization is softer, in contrast with our findings, that exhibit an edge in $P_z$ when the canted ferrimagnet transforms into the canted antiferromagnetic phase.
A possible solution to these anomalies can be the presence of a small orthogonal field. In Fig. \ref{fig:h110_Pz_b} we plotted our results for the same setting as in Fig. \ref{fig:h110_Pz_a}, but with a small applied field in $[\overline{1}10]$. Naturally in this case the twofold degeneracy of the canted ferrimagnetic phase is lifted as the perpendicular field selects between $\pm m_{[\overline{1}10]}$.

\begin{figure}[h]
\begin{center}
\includegraphics[width=8cm]{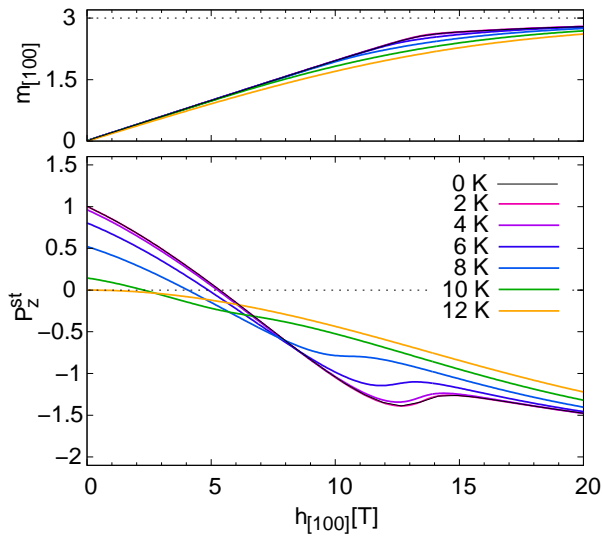}
\caption{Staggered polarization and magnetization results at finite temperature. We predict that in this rotated field setting the canted ferrimagnetic phase is missing, and the lower polarization curve is selected when $K_z>0$. Similarly to the $h||[110]$ case, the polarization depends strongly on the temperature, however the magnetization hardly changes.($J=J_z=1.885$ K, $\Lambda=15.08$ K, $D_z=-0.02$ K, $K_z=0.01$ K and $g=2.2$)}
\label{fig:h100_Pz}
\end{center}
\end{figure}

In Fig. \ref{fig:h100_Pz} we show the results for $h||[1,0,0]$. As mentioned before the relevant quantity in this case is the staggered polarization $P^{\text{st}}_z$. In good agreement with our zero temperature findings, the canted ferrimagnetic phase is absent, and below $h_c$ the canted antiferromagnetic phase takes place. Similarly to $h||[110]$ at higher fields we reach the partially magnetized uniform phase.


\section{Conclusion}
We investigated the field and temperature dependence of the induced polarization in the multiferroic compound Ba$_2$CoGe$_2$O$_7$. A detailed analysis has been given on the effect of DM interaction and an additional polarization-polarization term for two orientations of magnetic fields: $h||[110]$ and $h||[100]$. We found that in the former case an antiferro polarization coupling leads to the emergence of a canted ferrimagnetic phase when $0<h<h_{c2}\approx 1$ T. In this region the polarization decreases sharply reproducing qualitatively well the experimental findings in Ref.~[\onlinecite{Murakawa2010}]. Above $h_{c2}$ the canted antiferromagnetic phase takes place, in wich the spins rotate according to the increasing field so that at $h_c\approx 13$ T they are aligned and the uniform, partially magnetized phase appears. For $h||[100]$, however, the canted ferrimagnetic phase is absent, and at finite field only the canted antiferromagnetic and partially magnetized phases can be observed. Changing the polarization-polarization coupling to ferro type, the situation would be reversed: in $h||[110]$ there would be no sign of the canted ferrimagnet, while it would appear for $h||[100]$. Using the model that qualitatively recovers the $[110]$-field measurements, we gave a prediction to the field dependence of polarization when $h||[100]$. Applying finite temperature mean field theory we determined the polarization for both orientation of magnetic field at various temperatures. 
The mean field results capture qualitatively well the temperature dependence of polarization and magnetisation: while the former is very sensitive to the temperature change, the latter is almost unaffected by it. Based on our calculations, we believe that relevant information regarding the low field phase could be obtained from torque measurements. Furthermore, extending the magnetization measurements to higher fields can provide information about the partially polarized uniform phase.

\begin{acknowledgments}
We are pleased to thank H.~Murakawa for sending us the magnetization and polarization measurement data.
We also thank S.~Bord\'acs, T. Feh\'er,  A. J\'anossy, and I. K\'ezsm\'arki for stimulating discussions. This work was supported by Hungarian OTKA Grant Nos. K73455 and NN76727.
\end{acknowledgments}

\appendix


\bibliographystyle{unsrt}
\bibliographystyle{apsrev4-1}
\bibliography{BCGO_multif}
\end{document}